\documentclass[12pt,a4paper,titlepage,oneside]{article}

\usepackage[cp1252]{inputenc}
 
\usepackage[english]{babel}
\usepackage{cite, rotate}
\usepackage{amssymb}
\usepackage{amsmath}
\usepackage{epsfig,multicol}
\usepackage[bf]{subfigure}

\begin{document}

{\Large \bf Quaternionic Construction of the $W(F_{4} )$ Polytopes with Their Dual Polytopes and Branching under the Subgroups $W(B_{4} )$ and $W(B_{3} )\times W(A_{1} )$}
\\ 

{\bf Mehmet Koca\footnote{kocam@squ.edu.om}}, {\bf Mudhahir Al-Ajmi\footnote{mudhahir@squ.edu.om}} and {\bf Nazife Ozdes Koca\footnote{nazife@squ.edu.om}}

\noindent Department of Physics, College of Science, Sultan Qaboos University

\noindent P. O. Box 36, Al-Khoud 123, Muscat, Sultanate of Oman. \\


\section*{Abstract}4-dimensional $F_{4} $ polytopes and their dual
polytopes have been constructed as the orbits of the Coxeter-Weyl
group \textbf{$W(F_{4} )$} where the group elements and the vertices
of the polytopes are represented by quaternions. Branchings of an
arbitrary \textbf{$W(F_{4} )$} orbit under the Coxeter groups
\textbf{$W(B_{4} )$ } and \textbf{$W(B_{3} ) \times W(A_{1})$} have
been presented. The role of group theoretical technique and the use of
quaternions have been emphasized.

\section{Introduction}

\noindent Exceptional Lie groups $G_{2} {\rm [1],}F_{4} {\rm [2],}E_{6}
{\rm [3],}E_{7} {\rm [4]\; and\; }E_{8} {\rm [5]}$ have been proposed
as models in high energy physics. In particular, the largest
exceptional group $E_{8} $ turned out to be the unique gauge
symmetry $E_{8} \times E_{8} $ of the heterotic superstring theory [6].
It is not yet clear as to how any one of these groups will describe
the symmetry of any natural phenomenon. A recent experiment by Coldea
et.al [7] on neutron scattering over $CoNb_{2} O_{6}$ (cobalt niobate)
forming a one-dimensional quantum chain reveals evidence of the scalar
particles describable by the $E_{8}$~[8] symmetry. The Coxeter-Weyl
groups associated with exceptional Lie groups describe the symmetries
of certain polytopes. For example, the Coxeter-Weyl groups $W(E_{6}
),W(E_{7} ){\rm \; and\; }W(E_{8} )$ describe the symmetries of the
Gosset's polytopes [9] in six, seven and eight dimensions
respectively. The Coxeter-Weyl group $W(F_{4} )$ is a unique group in
the sense that it describes the symmetry of a uniform polytope 24-cell
in 4-dimensions which has no correspondence in any other dimensions.
It is also interesting from the point of view that the automorphism
group $Aut(F_{4} )\approx W(F_{4} ):C_{2}$ can be described by the use
of quaternionic representation of the binary octahedral group [10]
that will be described in the next section.

\noindent In this paper we study regular and semi-regular 4D
polytopes whose symmetries and vertices are described by the
quaternionic representations of the $W(F_{4} )$ symmetry and the
discrete quaternions respectively. We also give the projections of the
polytopes under the symmetries $W(B_{4})$ and $W(B_{3}
)\times W(A_{1} )$. The latter decomposition helps how to view 4D
polytopes in three dimensions. The dual polytopes of the group
$W(F_{4} )$ have not been constructed in any mathematical literature.
By following the method which we have developed for the Catalan solids~
[11], duals of the Archimedean solids, we construct the duals of the
regular and semi-regular $F_{4}$-polytopes.

\noindent The paper is organized as follows. In Section~2 we construct
the group $Aut(F_{4} )\approx W(F_{4} ):C_{2}$ in terms of the
quaternionic elements of the binary octahedral group [12]. A review of
construction of the maximal subgroups $W(B_{4})$ and $W(B_{3}
)\times W(A_{1} )$ in terms of quaternions is presented [13].
Section~3 deals with the regular and semi-regular $F_{4} $-polytopes
and their branchings under the maximal subgroups $W(B_{4} ){\rm \;
  and\; }W(B_{3} )\times W(A_{1} )$. In Section~4 we construct the
dual polytopes of the $F_{4} $-polytopes. Section~5 is devoted to
discussion and results.

\noindent 

\noindent 

\section{Construction of $Aut(F_{4} )\approx W(F_{4} ):C_{2}$ with
  quaternions}

In this section we introduce the quaternionic root system of the
Coxeter diagram of $F_{4} $ and give a proof that the $Aut(F_{4}
)\approx W(F_{4} ):C_{2}$ can be represented as the left-right action
of the quaternionic representation of the binary octahedral group. The
$F_{4}$ diagram with the quaternionic simple roots is shown in Figure~
1. Note that all simple roots of the Coxeter diagrams have equal
norms.

\begin{figure}
\begin{center}
  \centering \includegraphics[height=5cm]{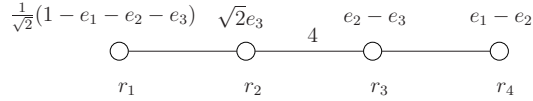} 
\caption{The $F_{4}$ diagram with quaternionic simple roots}
\end{center}
\end{figure}

\noindent Let $q=q_{0} e_{0} +q_{i} e_{i} $, $(e_{0} =1;i=1,2,3)$ be a
real unit quaternion with its conjugate defined by $\bar{q}=q_{0}
e_{0} -q_{i} e_{i} $ and the norm $q\bar{q}=\bar{q}q=1$ where the
quaternionic imaginary units satisfy the relations
\begin{equation} \label{GrindEQ__1_} e_{i} e_{i} =-\delta _{ij}
  +\varepsilon _{ijk} e_{k} , (i,j,k=1,2,3).
\end{equation} 
Here $\delta _{ij}$ and $\varepsilon _{ijk}$ are the Kronecker and
Levi-Civita symbols and summation over the repeated indices is
implied. We define the scalar product by the relation
\begin{equation} \label{GrindEQ__2_} (p,q)=\frac{1}{2}
  (\bar{p}q+\bar{q}p).
\end{equation} 
In general, a reflection generator $r$ of an arbitrary Coxeter group
with respect to a hyperplane represented by the vector $\alpha $ is
given by the action 
\begin{equation} \label{GrindEQ__3_}
  r:\Lambda \to \Lambda -\frac{2(\Lambda ,\alpha )}{(\alpha ,\alpha )}
  \alpha .
\end{equation} 
When $\Lambda $ and $\alpha$ are represented by quaternions the
equation \eqref{GrindEQ__3_} reads
\begin{equation} \label{GrindEQ__4_} r:\Lambda \to -\frac{\alpha
    \bar{\Lambda }\alpha }{(\alpha ,\alpha )} ,
\end{equation} 
where we define the product in \eqref{GrindEQ__4_} symbolically as
$r:\Lambda \to [-\frac{\alpha }{\sqrt{(\alpha ,\alpha )}}
,\frac{\alpha }{\sqrt{(\alpha ,\alpha )}}]^{*}
\Lambda $

\noindent and drop the factor $\Lambda$. Then the quaternionic
generators of the Coxeter group $W(F_{4} )$are given by
\begin{eqnarray} \label{GrindEQ__5_} r_{1} =[-\frac{1}{2} (1-e_{1}
  -e_{2} -e_{3} ),\frac{1}{2} (1-e_{1} -e_{2} -e_{3} )]^{*} ,r_{2}
  =[-e_{3} ,e_{3} ]^{*}, \nonumber \\ r_{3} =[-\frac{(e_{2} -e_{3}
    )}{\sqrt{2} } ,\frac{(e_{2} -e_{3} )}{\sqrt{2} } ]^{*} ,r_{4}
  =[-\frac{(e_{1} -e_{2} )}{\sqrt{2} } ,\frac{(e_{1} -e_{2}
    )}{\sqrt{2} } )]^{*} .
\end{eqnarray}

\noindent The Coxeter group $W(F_{4} )$ generated by these generators
can be compactly written as follows~[14]
\begin{equation} \label{GrindEQ__6_} W(F_{4} )=\{ [p,q]\oplus
  [r,s]\oplus [p,q]^{*} \oplus [r,s]^{*} \} ;{\rm \; \; \; \; }p,q\in
  T,{\rm \; \; }r,s\in T'.
\end{equation} 
Here the notation stands for $[p,q]\Lambda =p\Lambda q$ where $T{\rm \;
  and\; }T'$ are the subsets of the binary octahedral group $O=T\oplus
T'$ of order 48. They can further be written as the union of subsets
\begin{equation} \label{GrindEQ__7_} T=V_{0} \oplus V_{+} \oplus V_{-}
  ;{\rm \; \; \; }T'=V_{1} \oplus V_{2} \oplus V_{3}
\end{equation} 
where the subsets are given by the quaternions

\begin{eqnarray}
V_{0}&=&\{ \pm 1,\pm e_{1} ,\pm e_{2} ,\pm e_{3} \}, \nonumber \\  
V_{+}&=&\frac{1}{2} (\pm 1\pm e_{1} \pm e_{2} \pm e_{3}) {\rm \; (even\;
  number\; of\; }(+){\rm \; sign)\; }, \nonumber \\ 
V_{-}&=&\bar{V}_{+} \nonumber \\
  V_{1}&=&\{ {\tfrac{1}{\sqrt{2} }} (\pm 1\pm e_{1} ),{\tfrac{1}{\sqrt{2} }} (\pm e_{2} \pm e_{3} )\}, \nonumber \\   V_{2} &=&\{ {\tfrac{1}{\sqrt{2} }} (\pm 1\pm e_{2} ),{\tfrac{1}{\sqrt{2} }} (\pm e_{3} \pm e_{1} )\},   \nonumber \\ 
  V_{3}&=&\{ {\tfrac{1}{\sqrt{2} }} (\pm 1\pm e_{3} ),{\tfrac{1}{\sqrt{2} }} (\pm e_{1} \pm e_{2} )\}.
\end{eqnarray}

\noindent The set $V_{0}$ is the quaternion group and the set $T$ is
the binary tetrahedral group.

\noindent The set $V_{0}$ is an invariant subgroup of the binary
octahedral group $O=T\oplus T'$. The left or right coset decomposition
of the binary octahedral group under the quaternion group can be
written as $O=\sum _{i=1}^{6}g_{i} V_{0}$ with $g_{i}$ taking the
values
\begin{eqnarray} \label{GrindEQ__9_} 1, \frac{1}{\sqrt{2}} (e_{1}
  -e_{2} ),\frac{1}{\sqrt{2} } (e_{2} -e_{3} ), \frac{1}{\sqrt{2}} (e_{3} -e_{1}), \nonumber \\
  \frac{1}{2} (1+e_{1} +e_{2} +e_{3} ),\frac{1}{2}
  (1-e_{1} -e_{2} -e_{3} ).
\end{eqnarray} 
\begin{table}
\begin{center}
\begin{tabular}{|c|c|c|c|c|c|c|} \hline
    & $V_{0}$ & $V_{+}$ & $V_{-}$ & $V_{1}$ & $V_{2}$ & $V_{3}$ \\ \hline 
    $V_{0}$ & $V_{0}$ & $V_{+}$ & $V_{-}$ & $V_{1}$ & $V_{2}$ & $V_{3}$ \\ \hline 
    $V_{+}$ & $V_{+}$ & $V_{-}$ & $V_{0}$ & $V_{3}$ & $V_{1}$ & $V_{2}$ \\ \hline 
    $V_{-}$ & $V_{-}$ & $V_{0}$ & $V_{+}$ & $V_{2}$ & $V_{3}$ & $V_{1}$ \\ \hline 
    $V_{1}$ & $V_{1}$ & $V_{2}$ & $V_{3}$ & $V_{0}$ & $V_{+}$ & $V_{-}$ \\ \hline 
    $V_{2}$ & $V_{2}$ & $V_{3}$ & $V_{1}$ & $V_{-}$ & $V_{0}$ & $V_{+}$ \\ \hline 
    $V_{3}$ & $V_{3}$ & $V_{1}$ & $V_{2}$ & $V_{+}$ & $V_{-}$ & $V_{0}$ \\ \hline 
\end{tabular}
\caption{Multiplication table of the binary octahedral group.}
\end{center}
\end{table}
The coset representatives in \eqref{GrindEQ__9_} form a group
isomorphic to the permutation group $S_{3}$. This is reflected in the
multiplication Table~1. \\
Of course, the binary octahedral group has a larger invariant subgroup
$T$ which follows from the multiplication $TT\subset T,{\rm \; \;
}TT'\subset T',{\rm \; \; }T'T\subset T',{\rm \; }T'T'\subset T$. Now
we can write the equation \eqref{GrindEQ__6_} in a symbolic form as
\begin{equation} \label{GrindEQ__10_} 
W(F_{4} )=\{ [T,T]\oplus [T',T']\oplus [T,T]^{*} \oplus [T',T']^{*} \}.  
\end{equation} 
It is a group of order $288\times 4=1152$ from which one can identify
many maximal subgroups of order 576 [14]. The diagram symmetry of
$F_{4} ,{\rm \; }\alpha _{1} \leftrightarrow \alpha _{4} ,\alpha _{2}
\leftrightarrow \alpha _{3}$ can be generated by 
$D=[-\frac{1}{\sqrt{2} } (e_{2} +e_{3} ),e_{2} ]$. The extension of
the Coxeter group by the generator $D$ leads to the automorphism group 
$Aut(F_{4} )\approx W(F_{4} ):C_{2}$ of order 2304 and can be
compactly written as~[14]
\begin{equation} \label{GrindEQ__11_} 
Aut(F_{4} )\approx W(F_{4} ):C_{2} =\{ [O,O]\oplus [O,O]^{*} \} . 
\end{equation} 
The Cartan matrix of the Coxeter group $F_{4}$ and its inverse are given respectively by the matrices
\begin{eqnarray} \label{GrindEQ__12_} C_{F_{4} }
  =\left[\begin{array}{cccc} {2} & {-1} & {0} & {0} \\ {-1} & {2} &
      {-\sqrt{2} } & {0} \\ {0} & {-\sqrt{2} } & {2} & {-1} \\ {0} &
      {0} & {-1} & {2} \end{array}\right],
\end{eqnarray}
\begin{eqnarray} \nonumber
(C_{F_{4} } )^{-1}=\left[\begin{array}{cccc} {2} & {3} & {2\sqrt{2} } & {\sqrt{2} } \\
      {3} & {6} & {4\sqrt{2} } & {2\sqrt{2} } \\ {2\sqrt{2} } &
      {4\sqrt{2} } & {6} & {3} \\ {\sqrt{2} } & {2\sqrt{2} } & {3} &
      {2} \end{array}\right]
\end{eqnarray}
where the simple roots and the basis vectors in the dual space satisfy
respectively the relations ${\rm (}\alpha _{i} ,\alpha _{j} )=C_{ij}
,{\rm \; \; }(\omega _{i} ,\omega _{j} )=(C^{-1} )_{ij} $. Here the
basis vectors of the dual space are defined through the relation ${\rm
  (}\alpha _{i} ,\omega _{j} )=\delta _{ij}$. The $F_{4}$-polytopes
can be generated by the orbit $W(F_{4} )\Lambda $ where the vector
$\Lambda =a_{1} \omega _{1} +a_{2} \omega _{2} +a_{3} \omega _{3}
+a_{4} \omega _{4}$ is defined in the dual space which can also be
represented as $\Lambda =(a_{1} ,a_{2} ,a_{3} ,a_{4} )$. The
components of the vector $\Lambda $ are real numbers $a_{i} \ge 0$,
(i=1,2,3,4). The maximal Coxeter subgroups of the group $W(F_{4})$
that we deal with are $W(B_{4} )$ and $W(B_{3} )\times W(A_{1} )$.
There are two subgroups $W(B_{3L} )=<r_{1} {\rm ,}r_{2} {\rm ,}r_{3}>$
and $W(B_{3R} )=< r_{2} {\rm ,}r_{3} {\rm ,}r_{4}>$ which are not
conjugates in the group $W(F_{4} )$ but they are conjugates in the
larger group $Aut(F_{4} )$. We use the notation $\prec r_{1} {\rm
  ,}r_{2} {\rm ,}r_{3} ,...,r_{n} \succ $ for the Coxeter group
generated by reflection generators $r_{i}$, $(i=1,2,3,...,n)$.  The
octahedral group $W(B_{3} )$ plays an important role when an arbitrary
polytope of the group $W(F_{4} )$ is projected into 3D. The
quaternionic representations of the groups $W(B_{3} )\times C_{2}$ are
given by
\begin{subequations}  \label{GrindEQ__13_} 
\begin{align}
W(B_{3R} )\times C_{2} &=[T,\pm \bar{T}]\oplus [T',\pm
\bar{T}']\oplus [T,\pm \bar{T}]^{*} \oplus [T',\pm \bar{T}']^{*} \label{subeqn-1} \\
W(B_{3L} )\times C_{2} &=\{ [T,\pm \bar{q}\bar{T}q]\oplus [T',\pm \bar{q}\bar{T}'q]\oplus [T,\pm q\bar{T}q]^{*} \oplus [T',\pm q\bar{T}'q]^{*} \}, \nonumber \\ 
q=\frac{1+e_{1} }{\sqrt{2} } \label{subeqn-2}.
\end{align}
\end{subequations}
\noindent Here $[T,\pm \bar{T}]$ or any other pair represents the
group element $[p,\pm \bar{p}],{\rm \; }p\in T$. The Coxeter group 
$W(B_{3} )$ is isomorphic to the octahedral group $S_{4} \times C_{2}
$ of order 48. Therefore the group $W(B_{3} )\times C_{2} $ of order 96
can be embedded in the group $W(F_{4} )$ in 12 different ways, in each
embedding, a quaternion with $\pm $sign left invariant. In equations
(13a) and (13b) the groups leave the quaternions $\pm 1$ and $\pm
(\frac{1+e_{1} }{\sqrt{2} } )$ invariant respectively. It is clear
from these arguments that while the conjugates of the group $W(B_{3R}
)\times C_{2}$ leaving the vectors of the set $T'$ invariant the
conjugates of the group $W(B_{3L} )\times C_{2} $ leave invariant the
vectors from $T$.

\section{The $W(F_{4} )$ polytopes and branching under the group $W(B_{4} )$}

\begin{figure}
\begin{center}
  \centering \includegraphics[height=6cm]{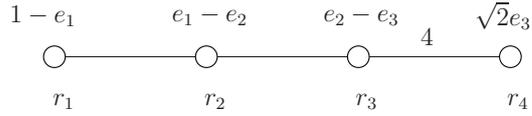} 
\caption{The $B_{4}$ diagram with quaternionic simple roots}
\end{center}
\end{figure}

\noindent The Coxeter diagram of the group $W(B_{4} )$ is shown in
Figure 2. The Cartan matrix and its inverse are respectively given by
the matrices

\noindent 
\begin{eqnarray} \nonumber 
C_{B_{4} }
  =\left[\begin{array}{cccc} {2} & {-1} & {0} & {0} \\ {-1} & {2} &
      {-1} & {0} \\ {0} & {-1} & {2} & {-\sqrt{2} } \\ {0} & {0} &
      {-\sqrt{2} } & {2} \end{array}\right],
\end{eqnarray} 

\begin{eqnarray} \label{GrindEQ__14_} 
(C_{B_{4} } )^{-1}
  =\left[\begin{array}{cccc} {1} & {1} & {1} & {\frac{1}{\sqrt{2} } }
      \\ {1} & {2} & {2} & {\sqrt{2} } \\ {1} & {2} & {3} &
      {\frac{3}{\sqrt{2} } } \\ {\frac{1}{\sqrt{2} } } & {\sqrt{2} } &
      {\frac{3}{\sqrt{2} } } & {2} \end{array}\right].
\end{eqnarray} 
If we chose the quaternionic simple roots as

\noindent 
\begin{equation} \label{GrindEQ__15_} 
\alpha _{1} =1-e_{1} ,{\rm \; \; }\alpha _{2} =e_{1} -e_{2} ,{\rm \; \; }\alpha _{3} =e_{2} -e_{3} ,{\rm \; \; }\alpha _{4} =\sqrt{2} e_{3}  
\end{equation}

\noindent the Coxeter group can be represented as~[10]

\noindent 
\begin{eqnarray} \label{GrindEQ__16_} W(B_{4} )&=& \{ [V_{0} ,V_{0}
  ]\oplus [V_{+} ,V_{-} ]\oplus [V_{-} ,V_{+} ]\oplus \nonumber \\& &
  [V_{1} ,V_{1} ]\oplus [V_{2} ,V_{2} ]\oplus [V_{3} ,V_{3} ]
  \nonumber \\& & \oplus [V_{0} ,V_{0} ]^{*} \oplus [V_{+} ,V_{-}
  ]^{*} \oplus [V_{-} ,V_{+} ]^{*} \nonumber \\& & \oplus [V_{1}
  ,V_{1} ]^{*} \oplus [V_{2} ,V_{2} ]^{*} \oplus [V_{3} ,V_{3} ]^{*}
  \}.
\end{eqnarray} 

\noindent The validity of the above equation can be checked by using
Table~1. The group $W(B_{4} )$ can be embedded in the group $W(F_{4}
)$ triply symmetric way which leads to the coset decomposition
\begin{eqnarray} \label{GrindEQ__17_}
&& W(F_{4})=\sum _{i=1}^{3}W(B_{4})g_{i}   \nonumber \\ 
&& g_{1}=[1,1],g_{2}=[\omega _{0} ,1],g_{3}=[\omega _{0} ^{2} ,1], \nonumber \\
&& \omega _{0}=\frac{1}{2} (1+e_{1} +e_{2}+e_{3} ).
\end{eqnarray}
When the vector $\Lambda =(a_{1} ,a_{2} ,a_{3} ,a_{4} )$ is expressed
in terms of quaternions we can find the vectors $\Lambda _{1} \equiv
\Lambda ,{\rm \; \; }\Lambda _{2} \equiv \omega _{0} \Lambda ,{\rm \;
  \; }\Lambda _{3} \equiv \omega _{0} ^{2} \Lambda $ and operate by
$W(B_{4} )$ on the left do determine the branching of the orbits of the
group $W(B_{4} )$ in the orbit of the group $W(F_{4} )\Lambda$. If all
components of the vector $\Lambda $ are different from zero then the
orbit $W(F_{4} )\Lambda $ represents a polytope with 1152 vertices. The
branching of the orbit $\Lambda =(a_{1} ,a_{2} ,a_{3} ,a_{4} )$ under
the orbits of the group $W(B_{4} )$ can be written as follows

\noindent 
\begin{eqnarray} \label{GrindEQ__18_} {(a_{1} ,a_{2} ,a_{3} ,a_{4}
    )_{F_{4}}} &=&\{ (\sqrt{2} a_{1} +\sqrt{2} a_{2} +a_{3} ),a_{4}
  ,a_{3} ,a_{2} \}_{B_{4}} \nonumber \\ & & +\{ a_{3} ,a_{4}
  ,(\sqrt{2} a_{2} +a_{3} ),a_{1} \}_{B_{4}} \nonumber \\ & & + \{
  (\sqrt{2} a_{2} +a_{3} ),a_{4} ,a_{3} ,(a_{1} +a_{2} ) \}_{B_{4}}.
\end{eqnarray}

\noindent We use the notation $W(G)\Lambda =(a_{1} ,a_{2} ,a_{3}
,a_{4} )_{G}$ for the orbit of the Coxeter group
where $G$ represents the Coxeter diagram. Let us give a few examples.
A complete list of decomposition of the regular and semi-regular
polytopes are given in the Appendix~1.

\noindent 

\noindent The polytope $(1,0,0,0)_{F_{4} } =W(F_{4} )\omega _{1}
=W(F_{4} )\sqrt{2} =\sqrt{2} T$ represents the set of 24 quaternions
$T=V_{0} \oplus V_{+} \oplus V_{-} $ up to a scale factor $\sqrt{2} $.
They are the elements of the the binary tetrahedral group and
represents the vertices of the regular polytope 24-cell. This is a
unique polytope which has no correspondence in any other dimensions.
This is perhaps due to the fact that $W(F_{4} )$ is associated with
the exceptional Lie group $F_{4} $. The polytope 24-cell consists of 24
octahedral cells, every vertex of which, is shared by 6 octahedra. To
illustrate this let us consider the vertex represented by the
quaternion 1. The following quaternions, e.g., represent the vertices
of an octahedron involving the quaternion 1:

\noindent 
\begin{equation} \label{GrindEQ__19_} 
1,{\rm \; }e_{1} ,{\rm \; }\frac{1}{2} (1+e_{1} \pm e_{2} \pm e_{3} ).                                                                        
\end{equation} 
The center of this octahedron is represented, up to a scale factor, by
the quaternion $\frac{1+e_{1} }{\sqrt{2} } \in T'$. If we introduce
the orthogonal vectors $p_{0} =\frac{1+e_{1} }{\sqrt{2} } ,p_{i}
=e_{i} p_{0} ,i=1,2,3$ then \eqref{GrindEQ__19_} can be written as
$p_{0} \pm p_{i}$ ,$(i=1,2,3)$ which represent an octahedron shifted
in the 4${}^{th}$ dimension by $p_{0} .$ We tabulate in equation
\eqref{GrindEQ__20_} the vertices of the six octahedra and their
centers sharing the vertex~1.
\begin{equation}\label{GrindEQ__20_} 
\begin{tabular}{cc}
{\rm Vertices} &  {\rm Centers(scaled)}    \\
1,$~e_{1}$,~$\frac{1}{2} (1+e_{1} \pm e_{2} \pm e_{3})$ & $1+e_{1}$  \\
1,$-e_{1}$,~$\frac{1}{2} (1-e_{1} \pm e_{2} \pm e_{3})$ & $1-e_{1}$  \\
1,$~e_{2}$,~$\frac{1}{2} (1 \pm e_{1} + e_{2} \pm e_{3})$ & $1+e_{2}$  \\
1,$-e_{2}$,~$\frac{1}{2} (1\pm e_{1} -e_{2} \pm e_{3})$ & $1-e_{2}$  \\
1,$~e_{3}$,~$\frac{1}{2} (1\pm e_{1}\pm e_{2} + e_{3})$ & $1+e_{3}$  \\
1,$-e_{3}$,~$\frac{1}{2} (1\pm e_{1} \pm e_{2} - e_{3})$ & $1-e_{3}$.  
\end{tabular}
\end{equation}
When we discuss the dual polytope of the $(1,0,0,0)_{F_{4} } $ we will
see that those centers of the six octahedra form another octahedron
with the vertices $1\pm e_{i} ,(i=1,2,3)$ where the vertices belong to
the set $T'=\frac{1}{\sqrt{2} } (0,0,0,1)_{F_{4} }$ which forms
another 24-cell. Actually the diagram symmetry operator $D$ exchanges
the two 24-cells so that the polytope 24-cell is said to be self-dual.

\noindent 

\noindent Now we can use \eqref{GrindEQ__18_} to check how 24-cell
decomposes under the subgroup $W(B_{4} )$. It is straightforward to
see that \\

\begin{equation} \label{GrindEQ__21_} 
  \begin{array}{l} {(1,0,0,0)_{F_{4} } =(\sqrt{2} ,0,0,0)_{B_{4} } +(0,0,0,1)_{B_{4} } .} \\ {{\rm \; \; \; \; \; \; \; \; \; \; \; \; \; \; \; \; \; \; \; \; \; \; \; \; \; \; \; \; \; }} \end{array} 
\end{equation} 
This is nothing but writing the set $T$ in the form $T=V_{0} \oplus
(V_{+} \oplus V_{-} )$. The set $V_{0} $ represents the vertices of a
hyperoctahedron (octahedron in 4D or called 16 cell) whose cells are
tetrahedra, 8 of which, share the same vertex. The set $(V_{+} \oplus
V_{-} )$ represents the hypercube (8-cell) with 16 vertices. The
hypercube consists of 8-cubes, 4 of which, meet at one vertex. The
hyperoctahedron and the hypercube are the duals of each other; the
centers of the 16 tetrahedra are represented by the vertices of the
hypercube and vice versa.

\noindent 

\noindent The dual polytope 24-cell $(0,0,0,1)_{F_{4} } $ decomposes
under the subgroup $W(B_{4} )$ as

\noindent 
\begin{equation} \label{GrindEQ__22_} 
(0,0,0,1)_{F_{4} } =(0,1,0,0)_{B_{4} }  
\end{equation} 
where the vertices, edges, faces and cells respectively decompose
as ${\rm 24=24}$, 96=96, 96=32+64, 24=16+8.

\noindent 

\section{The $W(F_{4} )$ polytopes and branching under the group
  $W(B_{3R} )$}

\noindent As we mentioned earlier there are two non-conjugate
octahedral groups in the Coxeter group $W(F_{4} )$. Here we will give
the projections of the $W(F_{4} )$ polytopes under the octahedral group
$W(B_{3R} )$. Decomposition with respect to the other group $W(B_{3L}
)$ can be done using a similar technique. The group $W(B_{3R} )$ up to
a conjugation can be represented by the quaternions as follows \\
\noindent 
\begin{equation} \label{GrindEQ__23_} W(B_{3R} )=\{ [T,\bar{T}]\oplus
  [T',\bar{T}']\oplus [T,\bar{T}]^{*} \oplus [T',\bar{T}']^{*} \}
\end{equation} 
which leaves the unit quaternion 1 invariant. The right coset
decomposition of the Coxeter group $W(F_{4} )$ under the octahedral
group $W(B_{3R} )$ can be given as \noindent $W(F_{4} )=\sum
_{i=1}^{24}W(B_{3R} )g_{i} {\rm \; with\; } g_{i} \in [T,1].$ Note
that $[T,1]{\rm \; and\; }[1,T]$ are two invariant subgroups of the
Coxeter group $W(F_{4} )$. Applying the group $W(F_{4} )$ on the
vector $\Lambda $ we obtain the orbit $W(F_{4} )\Lambda =\sum
_{i=1}^{24}W(B_{3R} )g_{i} \Lambda $. Defining the vectors $\Lambda
_{i} =g_{i} \Lambda ,{\rm \; }i=1,2,...,24{\rm \; with\; }\Lambda _{1}
\equiv \Lambda $ we obtain, in general, 24 different orbits of the
group $W(B_{3R} )$ in the decomposition of the orbit $W(F_{4} )\Lambda
$. Let a vector represented in the dual space of the root system of
the Coxeter group $W(B_{3R} )$ be given by $b_{1} v_{1} +b_{2} v_{2}
+b_{3} v_{3} $. When the quaternionic simple roots of the group
$W(B_{3R} )$ are given by $\alpha _{1} =\sqrt{2} e_{3} ,{\rm \;
}\alpha _{2} =e_{2} -e_{3} ,{\rm \; }\alpha _{3} =e_{1} -e_{2}$ then
basis vectors in the dual space are given by $v_{1} =\frac{1}{\sqrt{2}
} (e_{1} +e_{2} +e_{3} ),{\rm \; }v_{2} =e_{1} +e_{2} ,{\rm \; }v_{3}
=e_{1} .$ Applying the group $W(B_{3R} )$ on the quaternionic
representations of the vectors $\Lambda _{i} =g_{i} \Lambda ,{\rm \;
}i=1,2,...,24{\rm \; }$ and expressing them in the basis $v_{i}
, (i=1,2,3)$ then we obtain the following decomposition:
\begin{eqnarray} \label{GrindEQ__24_}
  (a_{1} ,a_{2} ,a_{3} ,a_{4})_{F_{4}} &=&\{ (a_{2} ,a_{3} ,a_{4} )_{B_{3} } \pm (a_{1} +\frac{3a_{2} }{2} +\frac{2a_{3} +a_{4} }{\sqrt{2} } )\} \nonumber \\
  &&+\{ (a_{2} ,a_{3} ,(\sqrt{2} a_{1} +\sqrt{2} a_{2} +a_{3} +a_{4} ))_{B_{3} }  \nonumber \\
  &&\pm (\frac{a_{2} }{2} +\frac{a_{3} +a_{4} }{\sqrt{2} } ) \}\nonumber \\
  &&+\{ (a_{2} ,(a_{3} +a_{4} ),(\sqrt{2} a_{1} +\sqrt{2} a_{2} +a_{3} ))_{B_{3} } \nonumber \\
  &&\pm (\frac{a_{2} }{2} +\frac{a_{3} }{\sqrt{2} } )\}  \nonumber \\
  &&+\{ ((a_{2} +\sqrt{2} a_{3} ),a_{4} ,(\sqrt{2} a_{1} +\sqrt{2} a_{2} +a_{3} ))_{B_{3} } \nonumber \\
  &&\pm (\frac{a_{2} }{2}) \} \nonumber \\
  &&+\{((a_{1} +2a_{2} +\sqrt{2} a_{3} ),a_{4} ,a_{3} )_{B_{3} } \nonumber \\ 
  &&\pm (\frac{a_{1}}{2}) \} \nonumber \\
  &&+\{((a_{1} +a_{2} ),a_{3} ,a_{4} )_{B_{3} } \nonumber \\
  &&\pm (\frac{1}{2} a_{1} +3a_{2} +2\sqrt{2} a_{3} +\sqrt{2} a_{4})\} \nonumber \\
  &&+ \{((a_{1} +a_{2} ),a_{3} ,(\sqrt{2} a_{2} +a_{3} +a_{4} ))_{B_{3} } \nonumber \\
  &&\pm \frac{1}{2}(a_{1} +a_{2} +\sqrt{2} a_{3} +\sqrt{2} a_{4})\} \nonumber \\
  &&+\{ ((a_{1} +a_{2} ),(a_{3} +a_{4} ),(\sqrt{2} a_{2} +a_{3} ))_{B_{3} } \nonumber \\
  &&\pm \frac{1}{2} (a_{1} +a_{2} +\sqrt{2} a_{3} )\} \nonumber \\
  &&+\{\frac{1}{2} (a_{1} +a_{2} +\sqrt{2} a_{3}), a_{4} ,(\sqrt{2} a_{2} +a_{3} ))_{B_{3} } \nonumber \\
  &&\pm \frac{1}{2} (a_{1} +a_{2}) \} \nonumber \\
  &&+ \{(a_{1} ,(\sqrt{2} a_{2} +a_{3} ),a_{4} )_{B_{3} } \nonumber \\
  &&\pm (\frac{a_{1} }{2} +a_{2} +\frac{2a_{3} +a_{4} }{\sqrt{2} }) \}  \nonumber \\
  &&+\{(a_{1} ,(\sqrt{2} a_{2} +a_{3} +a_{4} ),a_{3} )_{B_{3} } \nonumber \\
  &&\pm (\frac{a_{1} }{2} +a_{2} +\frac{a_{3} }{\sqrt{2} } )\}  \nonumber \\
  &&+\{ (a_{1} ,(\sqrt{2} a_{2} +a_{3} ),(a_{3} +a_{4} ))_{B_{3} } \nonumber \\
  &&\pm \frac{1}{2}(a_{1} +2a_{2} +\sqrt{2} a_{3} +\sqrt{2} a_{4} )\}.
\end{eqnarray}
\noindent The term starting with the $\pm $ represents a $W(A_{1}
)\approx C_{2} $ vector. For a general $W(F_{4} )$ orbit the formula
\eqref{GrindEQ__24_} shows the sphere $S^{3} $ representing the
vertices of the $W(F_{4} )$ polytope sliced by 24 hyperplanes. Each
crossection of the sphere $S^{3} $ with an hyperplane orthogonal to
the vector represented by the group $W(A_{1} )\approx C_{2} $ is an
$S^{2}$ sphere representing a particular $W(B_{3} )$ polyhedron. Let
us discuss a few cases. Consider the 24-cell represented by the orbit
$\frac{1}{\sqrt{2} } (1,0,0,0)_{F_{4} } =T$ which can be written as

\noindent 
\begin{eqnarray} \label{GrindEQ__25_}
  \frac{1}{\sqrt{2} } (1,0,0,0)_{F_{4} } &=&\{ (0,0,0)_{B_{3} } \pm (\frac{1}{\sqrt{2} }) \} +\{ (0,0,1)_{B_{3} } \pm (0)\}\nonumber \\
  && + \{ \frac{1}{\sqrt{2} } (1,0,0)_{B_{3} } \pm (\frac{1}{2\sqrt{2}
  } )\} .
\end{eqnarray}

\noindent When expressed in terms of quaternionic vertices they
correspond to the decomposition

\noindent 
\begin{equation} \label{GrindEQ__26_} T=\pm 1+(\pm e_{1} ,\pm e_{2}
  ,\pm e_{3} )+\frac{1}{2} (1\pm e_{1} \pm e_{2} \pm e_{3}
  )+\frac{1}{2} (-1\pm e_{1} \pm e_{2} \pm e_{3} ) .
\end{equation} 
Either from the analysis of the equation \eqref{GrindEQ__25_} [15] or
directly reading the equation \eqref{GrindEQ__26_} the points $\pm
1$ corresponds to the two opposite poles of the sphere $S^{3} $. The
set of vertices $(\pm e_{1} ,\pm e_{2} ,\pm e_{3} )$represent an
octahedron obtained as the intersection of the hyperplane through the
center of the sphere $S^{3} $. The last two sets of quaternions in
\eqref{GrindEQ__26_} represent cubes oppositely placed with respect to
the center of the sphere $S^{3} $. The dual 24-cell represented by the
orbit $\frac{1}{\sqrt{2} } (0,0,0,1)_{F_{4} } =T'$ decomposes as

\noindent 
\begin{equation} \label{GrindEQ__27_} \frac{1}{\sqrt{2} }
  (0,0,0,1)_{F_{4} } =\{ \frac{1}{\sqrt{2} } (0,0,1)_{B_{3} } \pm
  \frac{1}{2} \} +\{ \frac{1}{\sqrt{2} } (0,1,0)_{B_{3} } \pm (0)\}
\end{equation} 
which can also be represented by the set of quaternions
\begin{eqnarray} \label{GrindEQ__28_}
  T'&=&\{ \frac{1}{\sqrt{2} } (\pm 1\pm e_{1} ),\frac{1}{\sqrt{2} } (\pm 1\pm e_{2} ),\frac{1}{\sqrt{2} } (\pm 1\pm e_{3} )\} + \nonumber \\
  &&\{ \frac{1}{\sqrt{2} } (\pm e_{1} \pm e_{2} ),\frac{1}{\sqrt{2} }
  (\pm e_{2} \pm e_{3} ),\frac{1}{\sqrt{2} } (\pm e_{3} \pm e_{1} )\}
  .
\end{eqnarray} 
The first set represents two octahedra oppositely placed with respect
to the origin along the fourth direction, the second set represents a
cuboctahedron with 12 vertices around the center of the sphere $S^{3}
$. The decompositions of quaternionic vertices of 24- cell indicate
that it can have two different decompositions under the group
$W(B_{3R} )$. If the subgroup $W(B_{3L} )$ were chosen for the
projection of the 24-cell into 3D then the above decomposition of the
sets represented by $T{\rm \; and\; }T'$ would be interchanged. Using
equation \eqref{GrindEQ__24_}, any interested reader can work out the
projection of any $W(F_{4} )$ polytope into 3D with the octahedral
residual symmetry.

\noindent 

\section{Dual polytopes of the uniform polytopes of the Coxeter-Weyl group $W(F_{4} )$}

\noindent The Catalan solids which are the duals of the Archimedean
solids can be derived from the Coxeter diagrams $A_{3} ,{\rm \; }B_{3}
,{\rm \; }H_{3}$ with a simple technique [11]. Following the same
group theoretical technique we have already constructed the dual
polytopes of the $W(H_{4} )$ polytopes [16]. Here we apply the same
technique to determine the vertices and the symmetries of the dual
polytopes of the uniform \textbf{$W(F_{4} )$} polytopes. To find the
vertices of the dual polytope one should determine the vectors
representing the centers of the cells at the vertex $(a_{1} ,a_{2}
,a_{3} ,a_{4} )$. The relative magnitudes of the vectors of interest
can be obtained by noting that the hyperplane formed by the vectors
representing the centers of the cells are orthogonal to the
vertex $(a_{1} ,a_{2} ,a_{3} ,a_{4} )$. The dual polytopes are cell
transitive while the polytopes are vertex transitive. In this section
we will construct the vertices of a dual polytope and determine its
cell structure. We shall also use the group theoretical technique to
determine the numbers of vertices $N_{0} $, edges $N_{1} $, faces
$N_{2} $ and the cells $N_{3} $ of an arbitrary polytope. All 4D
polytopes satisfy the Euler's topological formula $N_{0} -N_{1} +N_{2}
-N_{3} =0$. \textbf{}

\subsection{Dual polytope of the 24-cell $(1,0,0,0)_{F_{4}}
  =W(F_{4})\omega _{1}$}

\noindent 

\noindent The 24-cell has 24 vertices, 96 edges, 240 faces and
24-cells. First we discuss how these numbers follow from the group
theoretical calculations. We note first that the vertex represented by
the vector $\omega _{1}$ is left invariant under the subgroup
$W(B_{3R} )=<r_{2} ,r_{3} ,r_{4}>$ of order 48. The number
of vertices then is the number of left coset representatives under the
decomposition

\noindent $W(F_{4} )=\sum _{i}g_{i} W(B_{3R} )$ which is equal to the
index $\frac{\left|W(F_{4} )\right|}{\left|W(B_{3R} )\right|}
=\frac{1152}{48} =24$ . The number of edges can be determined as
follows. The difference of the vectors $\omega _{1} -r_{1} \omega _{1}
=\alpha _{1} $ or $-\alpha _{1} $ represents an edge. The set $(\pm
\alpha _{1} )$is left invariant by the group generated by $<r_{1}
,r_{3} ,r_{4}>$ of order 12 so that its index in the group
\textbf{$W(F_{4} )$ } equals 96. Similarly from the vector $(1,0)$ of
the subgroup \textbf{$W(A_{2} )$} one can generate an equilateral
triangle by the group \textbf{$W(A_{2} )=\prec r_{1} ,r_{2} \succ $ }
of order 6.  Therefore the group leaving the triangle invariant is the
subgroup generated by \textbf{$<r_{1} ,r_{2} ,r_{4}>=W(A_{2} )\times
  W(A_{1} )$} of order 12 and its index in the group \textbf{$W(F_{4}
  )$} , 96, is the number of triangular faces. The octahedral cell is
generated by the group $W(B_{3L} )=<r_{1} ,r_{2} ,r_{3}>$ whose index,
24, is the number of octahedral cells hence the name of the polytope
24-cell.

\noindent The center of the octahedron $W(B_{3L} )\omega _{1} $ can be
represented by the vector $\omega _{4}$ up to a scale factor. The
number of octahedra at the vertex $\omega _{1} $ is given by the
formula

\noindent 
\begin{equation} \label{GrindEQ__29_} 
\frac{\left|<r_{2} ,r_{3,} r_{4}> \right|}{\left|< r_{2} ,r_{3}> \right|} =6 
\end{equation}

\noindent which is the index of the subgroup $<r_{2} ,r_{3}> $ of
order 8 fixing the vertex $\omega _{1}$ of the octahedron in the group
$<r_{2} ,r_{3} ,r_{4}>$ which fixes the vertex $\omega _{1} $. The
vertices of the cell representing the centers of the 6 octahedra can
be determined by applying $W(B_{3R} )=<r_{2} ,r_{3} ,r_{4}>$ on the
vector $\omega _{4}$ which leads to another octahedron. We have
discussed this case in detail in Section~3.

\subsection{Dual polytope of the polytope $(0,1,0,0)_{F_{4} } =W(F_{4} )\omega _{2}$}

\noindent Applying the technique discussed in Section 5.1 we determine
$N_{0} =96$ vertices, $N_{1} =288$ edges, $N_{2}=240$ faces and $N_{3}
=48$ cells. The polytope has 96 triangular, 144~square faces and 48
cells which consist of 24 cuboctahedra and 24 cubes whose symmetries
are respectively are the conjugate groups of the groups $W(B_{3L}
){\rm \; and\; }W(B_{3R} )$.

\noindent The number of cuboctahedral cells at the vertex $\omega _{2}
$ is determined by the index $\frac{\left|<r_{1} ,r_{3}
    ,r_{4}>\right|}{\left|<r_{1} ,r_{3} >\right|} =3$ . The group in
the numerator is a subgroup of the group \textbf{$W(F_{4} )$} fixing
the vertex $\omega _{2}$ and the group in the denominator is the
subgroup of the octahedral group $W(B_{3L} )$ fixing the same vertex.
The group $W(B_{3L} )$ which generates the cuboctahedra leaves the
vector $\omega _{4}$ invariant which can be taken as the vector
representing its center. So the vectors representing the centers of
the 3 cuboctahedra at the vertex $\omega _{2} $ can be taken as
$\omega _{4} ,{\rm \; }r_{3} r_{4} \omega _{4} ,{\rm \; (}r_{3} r_{4}
)^{2} \omega _{4} $. Similarly the number of the cubes at the same
vertex is 2 whose centers can be represented by the vectors $\lambda
\omega _{1} {\rm \; and\; }\lambda r_{1} \omega _{1} $ where$\lambda $
is a scale factor determined from the equation$ (\lambda \omega _{1}
{\rm -}\omega _{4} ).\omega _{2} =0{\rm \; as\; }\lambda
=\frac{2\sqrt{2} }{3} .$ These five vertices form a cell which is a
semi-regular bipyramid whose center is represented by the vector
$\omega _{2}$. The bipyramid consists of six isosceles triangles of
sides $\frac{\sqrt{10} }{3} {\rm \; and\; }\sqrt{2}$. It has the
symmetry $<r_{1} ,r_{3} ,r_{4}> \approx C_{2} \times D_{3}$
of order 12 which fixes the vector $\omega _{2}$. The set of vertices
of the dual polytope is the union of the orbits $\frac{2\sqrt{2} }{3}
(1,0,0,0)_{F_{4} } \oplus (0,0,0,1)_{F_{4} }$. They define two
concentric spheres $S^{3} $ with the fraction of radii $\frac{R_{4}
}{R_{1} } =\frac{3}{2\sqrt{2} } \approx 1.06$. Let us define the
quaternionic unit vectors $p_{0} =\frac{\omega _{2} }{\left|\omega
    _{2} \right|} =\frac{3+e_{1} +e_{2} +e_{3} }{2\sqrt{3} } ,{\rm \;
}p_{i} =e_{i} p_{0} ,{\rm \; }(i=1,2,3)$. Note that all the cells of the
dual polytope are bipyramids and their centers are the vertices of the
polytope $(0,1,0,0)_{F_{4} } .$ After expressing five vertices of the
bipyramid in terms of the new quaternionic units and omitting the
common 4${}^{th}$ component proportional to $p_{0} $, a scaled set of
the vectors representing the vertices of the bipyramid will read
\begin{eqnarray} \label{GrindEQ__30_} \omega _{4} \approx p_{1}
  -p_{3}, {\rm \; }r_{3} r_{4} \omega _{4} \approx -p_{1} +p_{2}, {\rm \; }r_{4} r_{3}
  \omega _{4} \approx -p_{2} +p_{3}, \nonumber \\ \lambda \omega _{1}
  \approx -\frac{2}{3} (p_{1} +p_{2} +p_{3} ),} {\rm \; }{\lambda r_{1} \omega
  _{1} \approx \frac{2}{3} (p_{1} +p_{2} +p_{3} ).
\end{eqnarray}

\noindent A plot of this bipyramid is shown in Figure 3.

\begin{figure}
\begin{center}
\subfigure[]{\epsfig{file=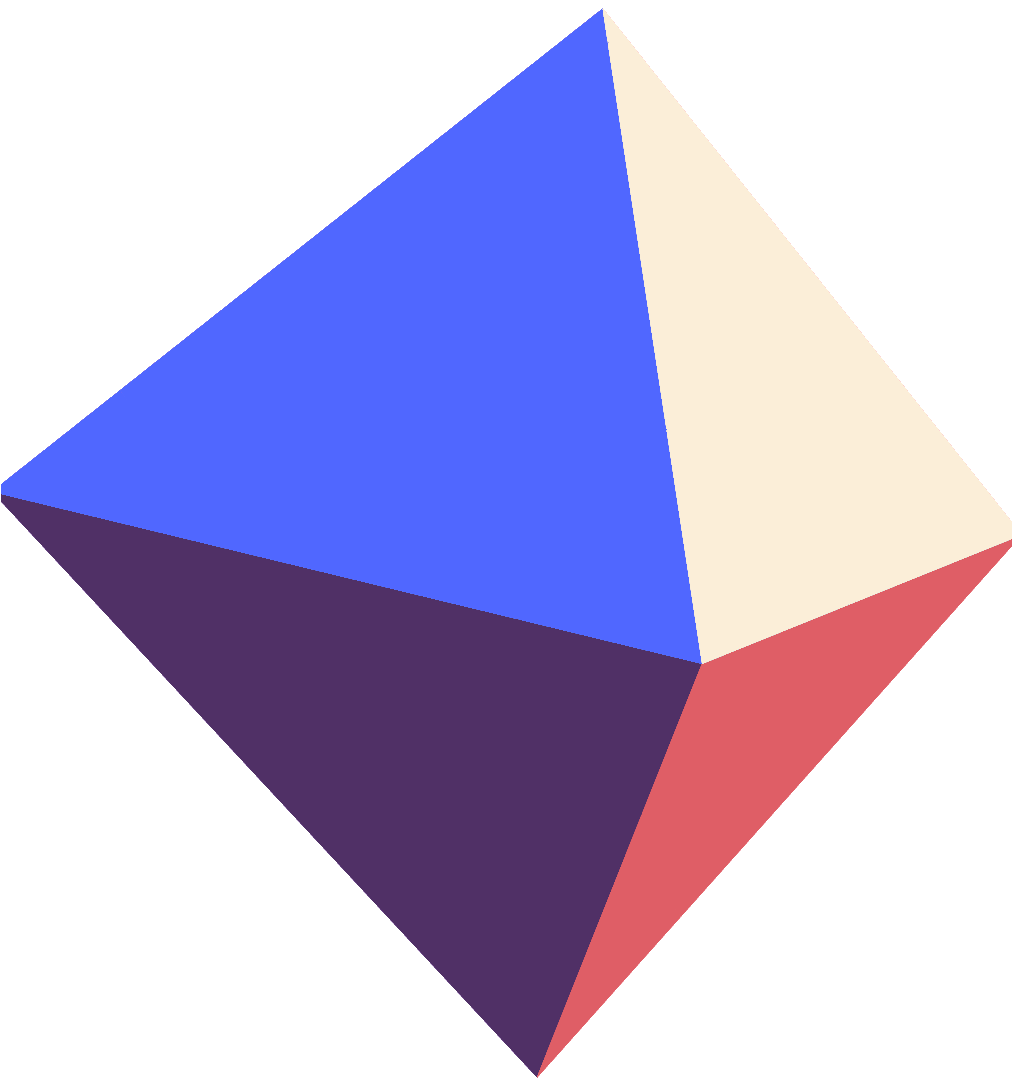,
      clip=,scale=0.4}}
\subfigure[]{\epsfig{file=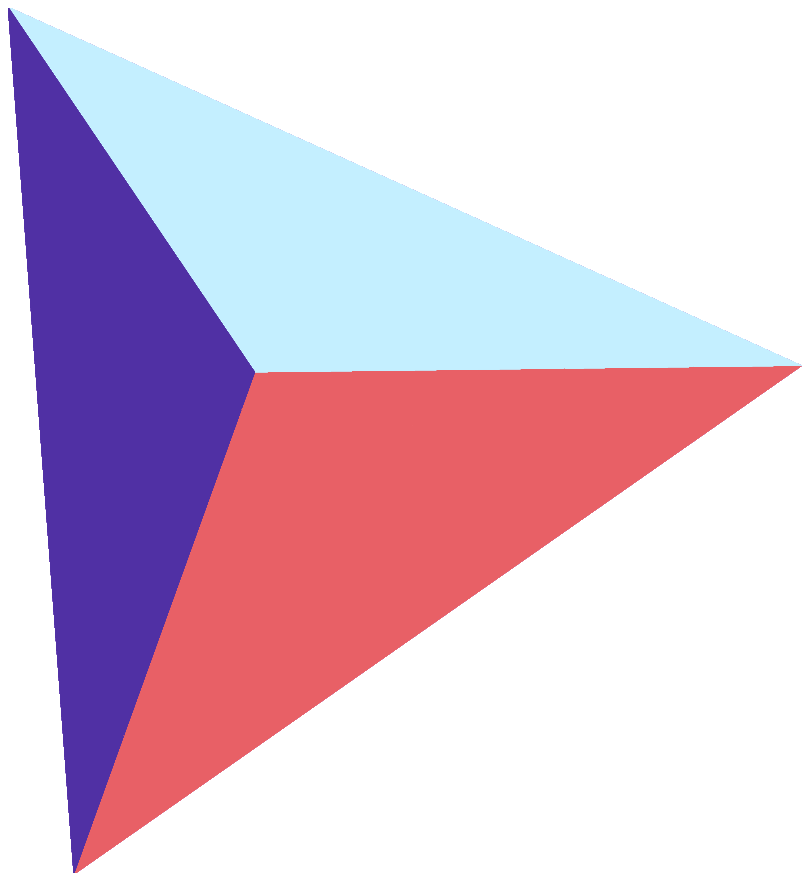,
      clip=,scale=0.5}}
\caption{Plot of the bipyramid representing the vertices in \eqref{GrindEQ__30_}. (a) Side view. (b) Top view.}
\end{center}
\end{figure}

\subsection{Dual polytope of the polytope $(0,0,1,0)_{F_{4} } =W(F_{4}
  )\omega _{3}$}

\noindent This polytope is obtained from the polytope
$(0,1,0,0)_{F_{4} } =W(F_{4} )\omega _{2} $ by exchanging $\omega _{2}
{\rm \; and\; }\omega _{3} .$

\subsection{Dual polytope of the polytope $(1,1,0,0)_{F_{4} } =W(F_{4}
  )(\omega _{1} +\omega _{2} )$}

\noindent This polytope has $N_{0} =192$ vertices, $N_{1} =384$ edges,
$N_{2} =240$ faces and $N_{3} =48$ cells. The faces are of two types:
96 hexagons and 144 squares. The cells consist of 24 truncated
octahedra and 24 cubes. To find the dual polytope we have to determine
the vectors representing the centers of these cells. The vectors
representing the centers of the truncated octahedra and the cube at
the vertex $(\omega _{1} +\omega _{2} )$

\noindent are given by
\begin{equation} \label{GrindEQ__31_} \omega _{4} ,{\rm \; }r_{3}
  r_{4} \omega _{4} ,{\rm \; (}r_{3} r_{4} )^{2} \omega _{4} ,{\rm \;
  }\lambda \omega _{1} .
\end{equation} 
The relative scale parameter is determined to be $\lambda
=\frac{3\sqrt{2} }{5} $ . The cell is a pyramid consisting of an
equilateral triangular base of side $\sqrt{2} $ and three isosceles
triangular faces of sides $\sqrt{2}$ and $\sqrt{\frac{26}{25} } .$
Defining the unit quaternion $p_{0} =\frac{\omega _{1} +\omega _{2}
}{\left|\omega _{1} +\omega _{2} \right|} =\frac{5+e_{1} +e_{2} +e_{3}
}{2\sqrt{7} } {\rm \; and\; }p_{i} =e_{i} p_{0} ,{\rm \; }(i=1,2,3)$ one
obtains the vertices in 3D of the cell of the dual polytope as

\noindent 
\begin{eqnarray} \label{GrindEQ__32_}
  \omega _{4} \approx 4p_{1} -2p_{3}, r_{3} r_{4} \omega _{4} \approx -2p_{2} +4p_{3}, \nonumber \\
  r_4 r_3 \omega _{4} \approx -2p_{1} +4p_{2}, \lambda \omega _{1} \approx
  -\frac{6}{5} (p_{1} +p_{2} +p_{3} ).
\end{eqnarray}
A plot of this cell is shown in Figure 4. It is clear that this solid
has a symmetry group generated by $<{\rm \; }r_{3} ,{\rm \;
}r_{4}>$.

\begin{figure}
\begin{center}
  \centering \includegraphics[height=5cm]{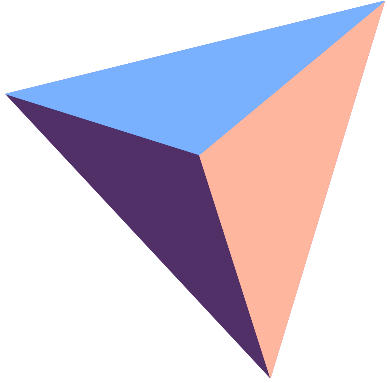} 
\caption{The plot of the solid with vertices given by \eqref{GrindEQ__32_}}
\end{center}
\end{figure}

\noindent The orbit of the vertices of the dual cell consists of two
concentric spheres $S^{3} $ with the ratio of the radii $\frac{R_{4}
}{R_{1} } =\approx 1.18.$

\noindent 

\subsection{Dual polytope of the polytope $(1,0,1,0)_{F_{4} } =W(F_{4}
  )(\omega _{1} +\omega _{3})$}

\noindent This polytope has $N_{0} =288$ vertices, $N_{1} =864$ edges,
$N_{2} =720$ faces, and $N_{3} =144$ cells. Reading from left to right
the faces consist of 96 triangles, 288+144 squares, and 192 triangles.
The cells consist of 24 small rhombicuboctahedra, 24 cubeoctahedra,
and 96 triangular prisms.  The vectors representing the centers of the
cells at the vertex $(\omega _{1} +\omega _{3} )$ are given by
\begin{equation} \label{GrindEQ__33_} \lambda \omega _{4} ,{\rm \;
  }\lambda r_{4} \omega _{4} ,{\rm \; }\rho \omega _{1} ,{\rm \;
  }\omega _{2} ,{\rm \; }r_{2} \omega _{2} .
\end{equation} 
The relative scale parameters are determined to be $\lambda
=\frac{1+9\sqrt{2} }{7} ,{\rm \; }\rho =\frac{5-\sqrt{2} }{2}$. The
cell of the dual polytope consists of five vertices. Defining the unit
quaternion $p_{0} =\frac{\omega _{1} +\omega _{3} }{\left|\omega _{1}
    +\omega _{3} \right|} =\frac{2+\sqrt{2} +e_{1} +e_{2}
}{\sqrt{8+2\sqrt{2} } } {\rm \; and\; }p_{i} =e_{i} p_{0} ,{\rm \;
}(i=1,2,3)$ one obtains the vertices in 3D of the cell of the dual
polytope as

\noindent 
\begin{eqnarray} \label{GrindEQ__34_}
  \lambda \omega _{4} &\approx& \frac{1+9\sqrt{2} }{7} [(1+\sqrt{2} )p_{1} -p_{2} -p_{3} ], \nonumber \\
  \lambda  r_{4} \omega _{4} &\approx& \frac{1+9\sqrt{2} }{7} [-p_{1}+(1+\sqrt{2} )p_{2} +p_{3} ], \nonumber \\
\rho \omega_{1} &\approx& \frac{2-5\sqrt{2} }{2} [p_{1} +p_{2}], \nonumber \\
\omega _{2} &\approx& [p_{1} +(1-\sqrt{2})p_{2} +(1+\sqrt{2} )p_{3}], \nonumber \\
r_{2} \omega _{2} &\approx& [(1-\sqrt{2} )p_{1} +p_{2}-(1+\sqrt{2})p_{3}].
\end{eqnarray} 

A plot of this cell is shown in Figure 5. It is clear that this solid
has a symmetry of the Klein four-group generated by $<{\rm \;
}r_{2} ,{\rm \; }r_{4}>$. The dual polytope consists of the
union of the orbits $\rho (1,0,0,0)_{F_{4} } \oplus (0,1,0,0)_{F_{4} }
\oplus \lambda (0,0,0,1)_{F_{4} } $ with 144 vertices, 288 cells , 864
faces and 720 edges. The radii of the $S^{3}$ spheres are respectively
$R_{1} \approx$ 2.536, $R_{2} \approx$ 2.450, $R_{4} \approx$ 2.774.

\begin{figure}
\begin{center}
  \centering \includegraphics[height=5cm]{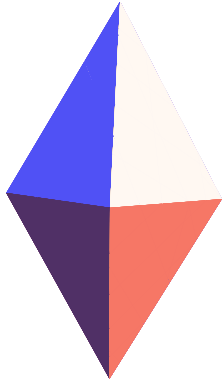} 
\caption{The plot of the solid with vertices given by
\eqref{GrindEQ__34_}}
\end{center}
\end{figure}

\noindent 

\subsection{Dual polytope of the polytope $(1,0,0,1)_{F_{4} } =W(F_{4}
  )(\omega _{1} +\omega _{4} )$}

The polytope has $N_{0}=144$ vertices, $N_{1}=576$ edges, $N_{2} =672$
faces and $N_{3}=240$ cells. Reading from left to right the faces
consist of 192 triangles, 288 squares, and 192 triangles. The cells
consist of 24+24 octahedra, 96+96 triangular prisms. The vectors
representing the centers of the cells at the vertex $(\omega _{1}
+\omega _{4} )$ are given by
\begin{eqnarray} \label{GrindEQ__35_}
  \lambda \omega _{1}, \lambda  \omega _{4}, \omega _{2}, r_{2} r_{3} \omega _{2}, (r_{2} r_{3} )^{2} \omega _{2}, (r_{2} r_{3} )^{3} \omega _{2}, \nonumber \\
\omega _{3}, r_{2} r_{3} \omega _{3}, (r_{2} r_{3} )^{2}
  \omega _{3}, (r_{2} r_{3})^{3} \omega _{3}.
\end{eqnarray}

\noindent The scale factor is found to be $\lambda =\frac{2+\sqrt{2}
}{2} .$ This cell of the dual polytope is a solid called tetragonal
trapezohedron with 10 vertices with the symmetry $D_{4} :C_{2} $ of
order 16. The dihedral group $D_{4} $ is generated by $D_{4} \approx
<r_{2} ,r_{3}>{\rm \; and\; }C_{2}$ is the group generated
by the diagram symmetry. Defining the new set of unit vectors by
$p_{0} =\frac{\omega _{1} +\omega _{4} }{\left|\omega _{1} +\omega
    _{4} \right|} $ ${\rm \; and\; }p_{i} =e_{i} p_{0} ,{\rm \;
}(i=1,2,3)$ we obtain the vertices of the solid in 3D as \\

\noindent 
\begin{equation} \label{GrindEQ__36_} 
  \begin{array}{l} {\lambda \omega _{1} \approx -p_{1} ,{\rm \; }\lambda \omega _{4} \approx p_{1} } \\ {\omega _{2} \approx (2\sqrt{2} -3)p_{1} +(\sqrt{2} -1)p_{2} +p_{3} } \\ {r_{2} r_{3} \omega _{2} \approx (2\sqrt{2} -3)p_{1} +p_{2} +(\sqrt{2} -1)p_{3} } \\ {(r_{2} r_{3} )^{2} \omega _{2} \approx (2\sqrt{2} -3)p_{1} +(\sqrt{2} -1)p_{2} -p_{3} } \\ {(r_{2} r_{3} )^{3} \omega _{2} \approx (2\sqrt{2} -3)p_{1} -p_{2} +(\sqrt{2} -1)p_{3} } \\ {\omega _{3} \approx (3-2\sqrt{2} )p_{1} +p_{2} +(\sqrt{2} -1)p_{3} } \\ {r_{2} r_{3} \omega _{3} \approx (3-2\sqrt{2} )p_{1} +(\sqrt{2} -1)p_{2} -p_{3} } \\ {(r_{2} r_{3} )^{2} \omega _{3} \approx (3-2\sqrt{2} )p_{1} -p_{2} -(\sqrt{2} -1)p_{3} } \\ {(r_{2} r_{3} )^{3} \omega _{3} \approx (3-2\sqrt{2} )p_{1} -(\sqrt{2} -1)p_{2} +p_{3}. } \end{array} 
\end{equation}

\begin{figure}
\begin{center}
\subfigure[]{\epsfig{file=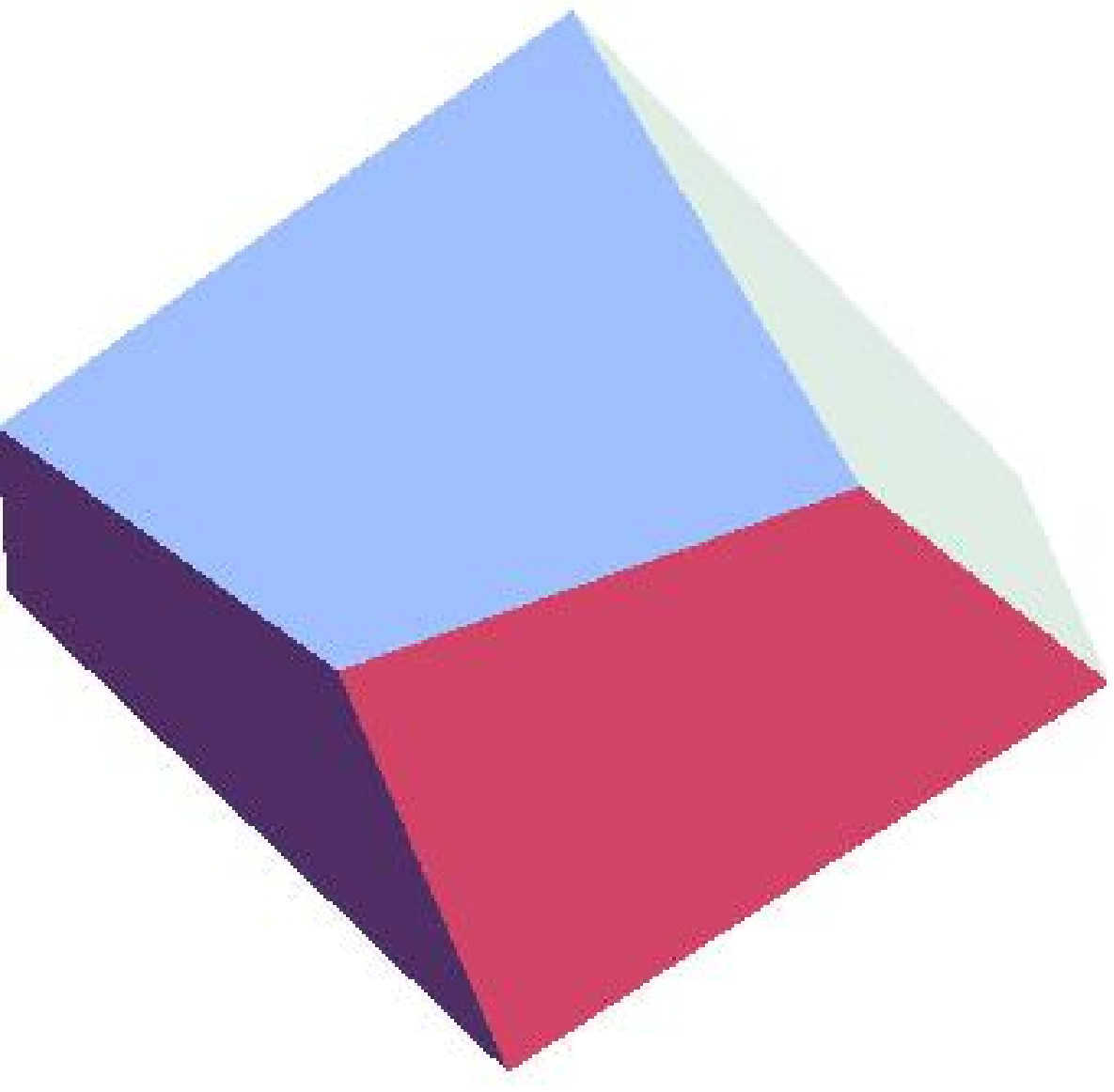,
      clip=,scale=0.5}}
\subfigure[]{\epsfig{file=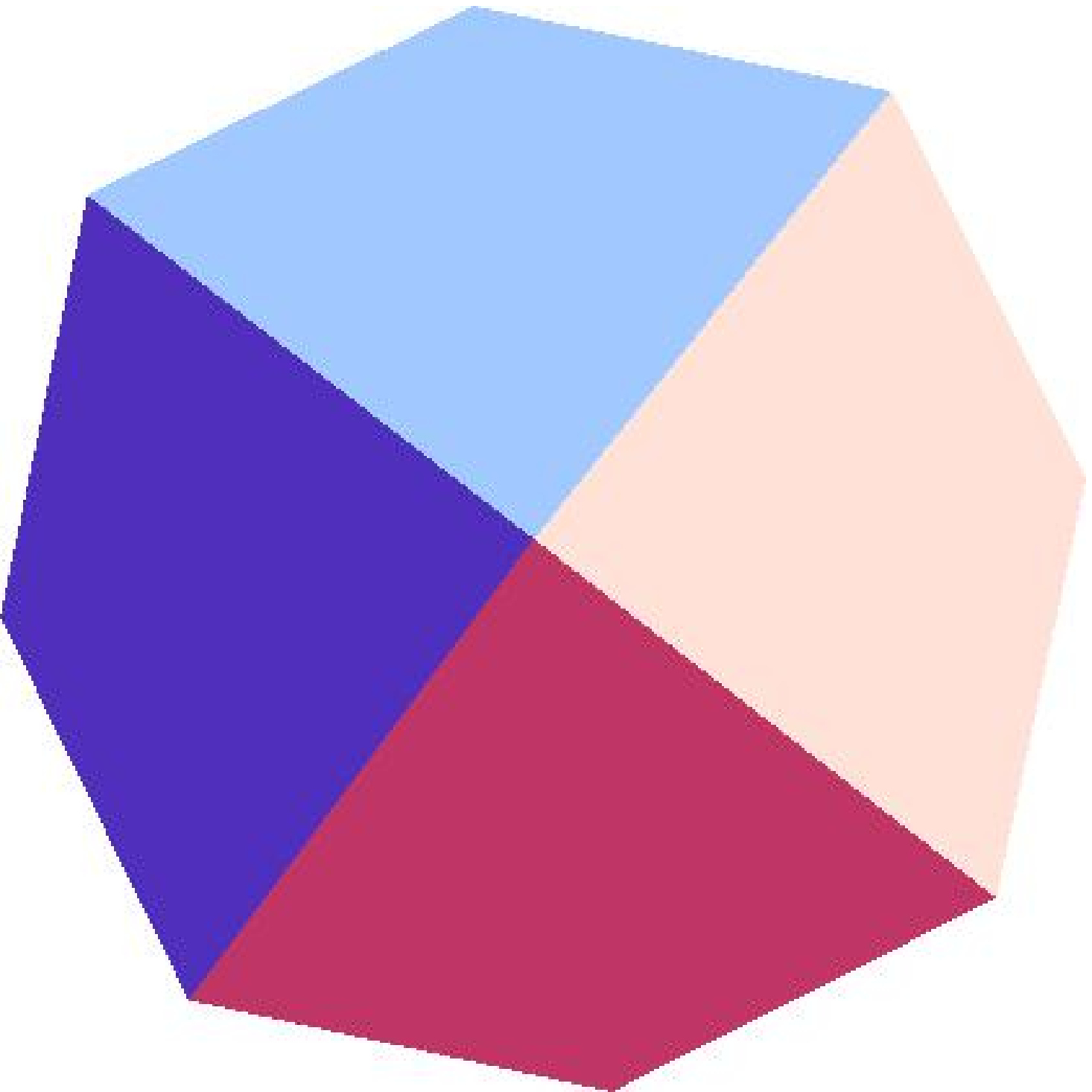,
      clip=,scale=0.4}}
  \caption{The plot of the tetragonal trapezohedron with vertices
    given by \eqref{GrindEQ__36_}. (a) Side view: Three kites meet at
    one vertex, (b) Top view: Four kites meet at one vertex.}
\end{center}
\end{figure}

\noindent A plot of this tetragonal trapezoid is shown in Figure 6.

It has eight faces and each face is a kite with sides
$\sqrt{16-10\sqrt{2} } \approx 1.363{\rm \; and\; }\sqrt{80-56\sqrt{2}
} \approx 0.897$ and the area 0.934.  So the dual polytope has 144
cells of tetragonal trapezohedra shown in Figure 6. The dual polytope
is cell transitive and it has 240 vertices 672 edges and 576 faces of
kites described above. The polytope is the union of the orbits
$\lambda (1,0,0,0)_{F_{4} } \oplus \lambda (0,0,0,1)_{F_{4} } \oplus
(0,1,0,0)_{F_{4} } \oplus (0,0,1,0)_{F_{4} } $. The vertices of the
dual polytope is on two concentric radii of the spheres $S^{3}$ with
radii $ R_{1}=R_{4}=$2.414 and $R_{2} =R_{3} =$2.449.

\subsection{Dual polytope of the polytope $(0,1,1,0)_{F_{4} } =W(F_{4}
  )(\omega _{2} +\omega _{3} )$}

The polytope has $N_{0} =288$ vertices, $N_{1} =576$ edges, $N_{2}
=336$ faces and $N_{3} =48$ cells. Reading from left to right the
faces consist of 96 triangles, 144 octagons, and 96 triangles. The
cells consist of 24+24 truncated cubes. The vectors representing the
centers of the cells at the vertex $(\omega _{2} +\omega _{3} )$ are
given by $\omega _{1} , r_{1} \omega _{1}, \omega _{4} , r_{4} \omega
_{4}$. It is a solid with four vertices which has a symmetry $D_{2}
:C_{2} $ of order 8. The dihedral group is generated by the generators
$D_{2} =<r_{1} ,r_{4}>$ and the group $C_{2}$ is generated
by the diagram symmetry given by $D=[-\frac{1}{\sqrt{2} } (e_{2}
+e_{3} ),e_{2} ]$. The cell of the dual polytope consists of four
faces of isosceles triangles of two sides $\sqrt{4-2\sqrt{2} }$ and
one side $\sqrt{2}$.  Defining the new set of unit vectors by $p_{0}
=\frac{\omega _{2} +\omega _{3} }{\left|\omega _{2} +\omega _{3}
  \right|} $${\rm \; and\; }p_{i} =e_{i} p_{0} ,{\rm \; }(i=1,2,3)$ we
obtain the vertices of the cell as
\begin{eqnarray} \label{GrindEQ__37_}
  \omega_{1} &\approx& -(1+\sqrt{2} )p_{1} -(1+\sqrt{2} )p_{2} -p_{3}, \nonumber \\
  r_{1} \omega_{1} &\approx& (1+\sqrt{2} )p_{1} +p_{2} +(1+\sqrt{2} )p_{3} \nonumber \\
  \omega_{4} &\approx& (1+\sqrt{2} )p_{1} -p_{2} -(1+\sqrt{2} )p_{3}, \nonumber \\
  r_{4} \omega_{4} &\approx& -(1+\sqrt{2} )p_{1} +(1+\sqrt{2} )p_{2}
  +p_{3}.
\end{eqnarray}

\noindent A plot is depicted in Figure 7. The dual polytope is the
union of the vertices determined by the polytopes $(1,0,0,0)_{F_{4} }
\oplus (0,0,0,1)_{F_{4} } .$ The 48 vertices of the dual polytope are
on the same $S^{3}$ sphere with radius $\sqrt{2} $. Of course it is
cell transitive just like any dual polytope is. In addition it is also
vertex transitive under the group $Aut(F_{4} ).$

\begin{figure}
\begin{center}
  \centering \includegraphics[height=5cm]{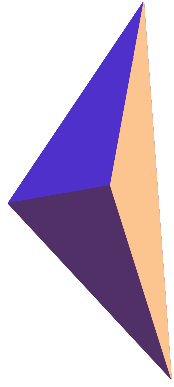} 
\caption{The plot of the cell of the dual polytope of the polytope $(0,1,1,0)_{F_{4} } =W(F_{4} )(\omega _{2} +\omega _{3} )$
}
\end{center}
\end{figure}

\subsection{Dual polytope of the polytope $(1,1,1,0)_{F_{4} }
  =W(F_{4} )(\omega _{1} +\omega _{2} +\omega _{3} )$}

\noindent The polytope has $N_{0} =576$ vertices, $N_{1} =1152$ edges,
$N_{2} =720$ faces and $N_{3} =144$ cells. Reading from left to right
the faces consist of 96 hexagons, 288 squares, 144 octagons and 192
triangles.  The cells consist of 24 great rhombicuboctahedra, 24
truncated cubes, and 96 triangular prisms. The vectors representing
the centers of the cells at the vertex $(\omega _{1} +\omega _{2}
+\omega _{3} )$ are given by ${\rm \; }\omega _{4} ,{\rm \; }r_{4}
\omega _{4} ,{\rm \; }\lambda \omega _{1} ,{\rm \; and\; }\rho \omega
_{2} .$ It is a solid with four vertices which has a symmetry
generated by $<r_{4}>$. The parameters are determined to
be $\lambda =\frac{3(1+3\sqrt{2} )}{17} ,{\rm \; }\rho
=\frac{3(1+\sqrt{2} )}{9+4\sqrt{2} } .$ The vertices of the dual cell
is the union of the orbits $\lambda (1,0,0,0)_{F_{4} } \oplus \rho
(0,1,0,0)_{F_{4} } \oplus (0,0,0,1)_{F_{4} } $. The vertices lie on
three $S^{3} $ spheres with the radii $R_{1} :R_{2} :R_{4}
=$1.31:1.21:1.41. Defining the new set of unit vectors by $p_{0}
=\frac{\omega _{1} +\omega _{2} +\omega _{3} }{\left|\omega _{1}
    +\omega _{2} +\omega _{3} \right|}$ ${\rm \; and\; }p_{i} =e_{i}
p_{0} ,{\rm \; }(i=1,2,3)$ we obtain the vertices of the cell as
\begin{eqnarray} \label{GrindEQ__38_}
  \lambda \omega _{1} &\approx& -\frac{3(8+7\sqrt{2} )}{17} p_{1} -\frac{3(8+7\sqrt{2} )}{17} p_{2} -\frac{3(6+\sqrt{2} )}{17} p_{3}, \nonumber \\
  \rho \omega _{2} &\approx& \frac{3}{9+4\sqrt{2} } [(2+\sqrt{2} )p_{1} +(2+\sqrt{2} )p_{2} +(4+3\sqrt{2} )p_{3}], \nonumber \\
\omega _{4} &\approx& (4+\sqrt{2} )p_{1} -\sqrt{2} p_{2} -(2+\sqrt{2} )p_{3}, \nonumber \\
r_{4} \omega _{4} &\approx& -(2+\sqrt{2} )p_{1} +(4+\sqrt{2} )p_{2}
+\sqrt{2} p_{3}.
\end{eqnarray}

\noindent A plot of this solid is depicted in Figure 8.

\begin{figure}
\begin{center}
  \centering \includegraphics[height=5cm]{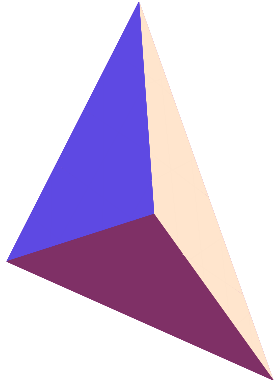} 
\caption{The plot of the cell of the dual polytope of the polytope $(1,1,1,0)_{F_{4} } =W(F_{4} )(\omega _{1} +\omega _{2} +\omega _{3} )$
}
\end{center}
\end{figure}

\subsection{Dual polytope of the polytope $(1,1,0,1)_{F_{4} } =W(F_{4}
  )(\omega _{1} +\omega _{2} +\omega _{4} )$}

The polytope has $N_{0} =576$ vertices, $N_{1} =1440$ edges, $N_{2}
=1104$ faces and $N_{3} =240$ cells. Reading from left to right the
faces consist of 192 hexagons, 144+288+288 squares, and 192 triangles.
The cells consist of 24 truncated octahedra, 24 small
rhombicuboctahedra, 96 hexagonal prisms, and 96 triangular prisms. The
vectors representing the centers of the cells at the vertex $(\omega
_{1} +\omega _{2} +\omega _{4} )$ are given by ${\rm \; }\omega _{3}
,{\rm \; }r_{3} \omega _{3} ,{\rm \; }\lambda \omega _{4} ,{\rm \;
  and\; }\rho \omega _{1} ,\eta \omega _{2} .$ It is a solid with five
vertices which has a symmetry generated by $<r_{3}>$. The
parameters are determined to be $\lambda
=\frac{3+6\sqrt{2}}{2+3\sqrt{2} } ,{\rm \; }\rho =\frac{3+6\sqrt{2}
}{5+\sqrt{2} } ,$ and $\eta =\frac{3+6\sqrt{2}}{9+2\sqrt{2} } .$ The
vertices of the dual cell is the union of the orbits $\rho
(1,0,0,0)_{F_{4} } \oplus \eta (0,1,0,0)_{F_{4} } \oplus
(0,0,1,0)\oplus \lambda (0,0,0,1)_{F_{4} } .$ Defining the new set of
unit vectors by $p_{0} =\frac{\omega _{1} +\omega _{2} +\omega _{4}
}{\left|\omega _{1} +\omega _{2} +\omega _{4} \right|} $ ${\rm \;
  and\; }p_{i} =e_{i} p_{0} ,{\rm \; }(i=1,2,3)$ we obtain the vertices
of the cell as
\begin{equation} \label{GrindEQ__39_} 
  \begin{array}{l} {\rho \omega _{1} \approx -\frac{3}{5+\sqrt{2} } [(6+5\sqrt{2} )p_{1} +(4+\sqrt{2} )p_{2} +(4+\sqrt{2} )p_{3} ],{\rm \; }} \\ {\eta \omega _{2} \approx \frac{6(1+2\sqrt{2} )}{4+9\sqrt{2} } [(1-\sqrt{2} )p_{1} +p_{2} +(1+\sqrt{2} )p_{3} ],} \\ {\lambda \omega _{4} \approx \frac{12(1+2\sqrt{2} )}{2+3\sqrt{2} } p_{1} -\frac{6(1+2\sqrt{2} )}{2+3\sqrt{2} } p_{3} ,} \\ {\omega _{3} \approx (2-\sqrt{2} )p_{1} +(4+\sqrt{2} )p_{2} +(-2+\sqrt{2} )p_{3} ,} \\ {r_{3} \omega _{3} \approx (4-\sqrt{2} )p_{1} -(2+\sqrt{2} )p_{2} +(2+\sqrt{2} )p_{3}. } \end{array} 
\end{equation} 

\noindent A plot of this solid is depicted in Figure 9.

\begin{figure}
\begin{center}
  \centering \includegraphics[height=5cm]{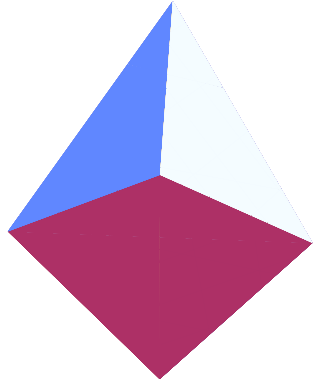} 
\caption{The plot of the cell of the dual polytope of the polytope $(1,1,0,1)_{F_{4} } =W(F_{4} )(\omega _{1} +\omega _{2} +\omega _{4} )$
}
\end{center}
\end{figure}

\subsection{Dual polytope of the polytope $(1,1,1,1)_{F_{4} } =W(F_{4}
  )(\omega _{1} +\omega _{2} +\omega _{3} +\omega _{4} )$}

The polytope has $N_{0} =1152$ vertices,$N_{1} =2304$ edges, $N_{2}
=1392$ faces and $N_{3} =240$ cells.  Reading from left to right the
faces consist of 192 hexagons, 288+288+288 squares, 144 octagons and
192 hexagons.  The cells consist of 24+24 great rhombicuboctahedra and
96+96 haxagonal prisms. The vectors representing the centers of the
cells at the vertex $(\omega _{1} +\omega _{2} +\omega _{3} +\omega
_{4})$ are given by $\omega _{1}, \omega _{4}$, and $\rho \omega _{2}
,\rho \omega _{3}$ with $\rho =\frac{5+3\sqrt{2} }{9+6\sqrt{2}}$. It
is a solid with four vertices which has a diagram symmetry only.  The
240 vertices of the dual cell is the union of the orbits
$(1,0,0,0)_{F_{4} } \oplus \rho (0,1,0,0)_{F_{4} } \oplus \rho
(0,0,1,0)\oplus (0,0,0,1)_{F_{4} } .$ Defining the new set of unit
vectors by $p_{0} =\frac{\omega _{1} +\omega _{2} +\omega _{3} +\omega
  _{4} }{\left|\omega _{1} +\omega _{2} +\omega _{3} +\omega _{4}
  \right|} $ ${\rm \; and\; }p_{i} =e_{i} p_{0} ,{\rm \; }(i=1,2,3)$ we
obtain the vertices of the cell as
\begin{equation} \label{GrindEQ__40_} 
  \begin{array}{l} {\omega _{1} \approx -(4+\sqrt{2} )p_{1} -(2+\sqrt{2} )p_{2} -\sqrt{2} p_{3} ,{\rm \; }} \\ {\omega _{4} \approx (4+\sqrt{2} )p_{1} -\sqrt{2} p_{2} -(2+\sqrt{2} )p_{3} ,} \\ {\rho \omega _{2} \approx \frac{5+3\sqrt{2} }{9+6\sqrt{2} } [(\sqrt{2} -2)p_{1} +(\sqrt{2} -2)p_{2} +(\sqrt{2} +4)p_{3} ],} \\ {\rho \omega _{3} \approx \frac{5+3\sqrt{2} }{9+6\sqrt{2} } [(-\sqrt{2} +2)p_{1} +(\sqrt{2} +4)p_{2} +(\sqrt{2} -2)p_{3} ].} \end{array} 
\end{equation} 

Note that the Coxeter diagram symmetry leads to the transformation
$p_{0} \to p_{0} ,{\rm \; }p_{1} \to -p_{1} ,{\rm \; }p_{2}
\leftrightarrow p_{3} {\rm \; which\; results\; in\; }\omega _{1}
\leftrightarrow \omega _{4} {\rm \; and\; }\rho \omega _{2}
\leftrightarrow \rho \omega _{3} .$ A plot of this solid is depicted
in Figure 10.

\begin{figure}
\begin{center}
  \centering \includegraphics[height=5cm]{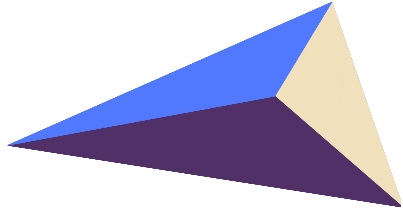} 
\caption{The plot of the cell of the dual polytope of the polytope
$(1,1,1,1)_{F_{4} } =W(F_{4} )(\omega _{1} +\omega _{2} +\omega _{3}
+\omega _{4} )$}
\end{center}
\end{figure}

\section{Conclusion}

4D polytopes can be classified with respect to their symmetries
represented by the Coxeter groups $W(A_{4} )$, $W(B_{4} )$, $W(H_{4}
)$ and $W(F_{4} )$. Here we constructed the regular and semiregular
polytopes of the Coxeter group $W(F_{4} )$ with their dual polytopes.
We have represented the group elements in terms of quaternions. It is
an interesting observation that the automorhism group of the Coxeter
diagram $F_{4}$ can be determined in terms of the binary octahedral
group as $Aut(F_{4} )=\{ [O,O]\oplus [O,O]^{*} \}$. The dual polytopes
of the $W(F_{4} )$ have been constructed for the first time in this
work.  We have explicitly given the decomposition of a given polytope
under the maximal subgroup $W(B_{4} )$. The projection of an arbitrary
$W(F_{4} )$ polytope in 3D space has been made with the use of the
octahedral group $W(B_{3} )$ which allows us to view the 4D polytope
in 3D space. For a detail decomposition see Appendix~2. This work can
be related to the quantum information theory based on 2-qubits since
we use a finite subgroup of the orthogonal group
$SU\eqref{GrindEQ__2_}\times
SU\eqref{GrindEQ__2_}\approx O\eqref{GrindEQ__4_}/C_{2} $. \\

\section*{Appendix 1}

The decompositon of the $W(F_4)$ wieghts under $W(B_4)$ is given as follows: \\

$(0,  0,  0,  1)_{F_4}$= $(0,1,0,0)_{B_4}$ \newline

$(0,  0,  1,  0)_{F_4}$= $(1,0,1,0)_{B_4}$ \newline

$(0,  0,  1,  1)_{F_4}$= $(1,1,1,0)_{B_4}$ \newline

$(0,  1,  0,  0)_{F_4}$= $(0,0,\sqrt{2},0)_{B_4}+(\sqrt{2},0,0,1)_{B_4}$. \newline

$(0,  1,  0,  1)_{F_4}$= $(0,1,\sqrt{2},0)_{B_4}+(\sqrt{2},1,0,1)_{B_4}$. \newline

$(0,  1,  1,  0)_{F_4}$= $(1,0,1+\sqrt{2},0)_{B_4}+(1+\sqrt{2},0,1,1)_{B_4}$. \newline

$(0,  1,  1,  1)_{F_4}$= $(1,1,1+\sqrt{2},0)_{B_4}+(1+\sqrt{2},1,1,1)_{B_4}$. \newline

$(1,  0,  0,  0)_{F_4}$= $(0,0,0,1)_{B_4}+(\sqrt{2},0,0,0)_{B_4}$. \newline

$(1,  0,  0,  1)_{F_4}$= $(0,1,0,1)_{B_4}+(\sqrt{2},1,0,0)_{B_4}$. \newline

$(1,  0,  1,  0)_{F_4}$= $(1,0,1,1)_{B_4}+(1+\sqrt{2},0,1,0)_{B_4}$. \newline

$(1,  0,  1,  1)_{F_4}$= $(1,1,1,1)_{B_4}+(1+\sqrt{2},1,1,0)_{B_4}$. \newline

$(1,  1,  0,  0)_{F_4}$= $(0,0,\sqrt{2},1)_{B_4}+(\sqrt{2},0,0,2)_{B_4}+(2 \sqrt{2},0,0,1)_{B_4}$. \newline

$(1,  1,  0,  1)_{F_4}$= $(0,1,\sqrt{2},1)_{B_4}+(\sqrt{2},1,0,2)_{B_4}+(2 \sqrt{2},1,0,1)_{B_4}$. \newline

$(1,  1,  1,  0)_{F_4}$= $(1,0,1+\sqrt{2},1)_{B_4}+(1+\sqrt{2},0,1,2)_{B_4}+(1+2 \sqrt{2},0,1,1)_{B_4}$. \newline

$(1,  1,  1,  1)_{F_4}$= $(1,1,1+\sqrt{2},1)_{B_4}+(1+\sqrt{2},1,1,2)_{B_4}+(1+2 \sqrt{2},1,1,1)_{B_4}$. \newline

\appendix
\section*{Appendix 2}

The decompositon of the $W(F_4)$ wieghts under $W(B_3)$ is given as follows: \\

\noindent
$(0,0,0,1)_{F_4}$= \newline
\noindent
\\$\{(0,0,1)_{B_3}\pm(\frac{1}{\sqrt{2}})\}$+$\{(0,1,0)_{B_3}\pm(0)\}$. \\
\noindent
\\ $(0,0,1,0)_{F_4}$= \newline
\noindent
\\$\{(0,1,0)_{B_3}\pm(\sqrt{2})\}$+$\{(0,1,1)_{B_3}\pm(\frac{1}{\sqrt{2}})\}$+$\{(\sqrt{2},0,1)_{B_3}\pm(0)\}$. \\
\noindent
\\ $(0,0,1,1)_{F_4}$= \newline
\noindent
\\$\{(0,1,1)_{B_3}\pm(\frac{3}{\sqrt{2}})\}$+$\{(0,1,2)_{B_3}\pm(\sqrt{2})\}$+$\{(0,2,1)_{B_3}\pm(\frac{1}{\sqrt{2}})\}$+$\{(\sqrt{2},1,1)_{B_3}\pm(0)\}$. \\
\noindent
\\ $(0,1,0,0)_{F_4}$= \newline
\noindent
\\$\{(0,\sqrt{2},0)_{B_3}\pm(1)\}$+$\{(1,0,0)_{B_3}\pm(\frac{3}{2})\}$+$\{(1,0,\sqrt{2})_{B_3}\pm(\frac{1}{2})\}$+$\{(2,0,0)_{B_3}\pm(0)\}$. \\ \newpage
\noindent
\\ $(0,1,0,1)_{F_4}$= \newline
\noindent
\\$\{(0,\sqrt{2},1)_{B_3}\pm(1+\frac{1}{\sqrt{2}})\}$+$\{(0,1+\sqrt{2},0)_{B_3}\pm(1)\}$+$\{(1,0,1)_{B_3}\pm(\frac{3}{2}+\frac{1}{\sqrt{2}})\}$+$\{(1,0,1+\sqrt{2})_{B_3}\pm(\frac{1}{2}+\frac{1}{\sqrt{2}})\}$+$\{(1,1,\sqrt{2})_{B_3}\pm(\frac{1}{2})\}$+$\{(2,1,0)_{B_3}\pm(0)\}$. \\
\noindent
\\ $(0,1,1,0)_{F_4}$= \newline
\noindent
\\$\{(0,1+\sqrt{2},0)_{B_3}\pm(1+\sqrt{2})\}$+$\{(0,1+\sqrt{2},1)_{B_3}\pm(1+\frac{1}{\sqrt{2}})\}$+$\{(1,1,0)_{B_3}\pm(\frac{3}{2}+\sqrt{2})\}$+$\{(1,1,1+\sqrt{2})_{B_3}\pm(\frac{1}{2}+\frac{1}{\sqrt{2}})\}$+$\{(1+\sqrt{2},0,1+\sqrt{2})_{B_3}\pm(\frac{1}{2})\}$+$\{(2+\sqrt{2},0,1)_{B_3}\pm(0)\}$. \newline   
\noindent
\\ $(0,1,1,1)_{F_4}$= \newline
\noindent
\\$\{(0,1+\sqrt{2},1)_{B_3}\pm(1+\frac{3}{\sqrt{2}})\}$+$\{(0,1+\sqrt{2},2)_{B_3}\pm(1+\sqrt{2})\}$+
$\{(0,2+\sqrt{2},1)_{B_3}\pm(1+\frac{1}{\sqrt{2}})\}$+$\{(1,1,1)_{B_3}\pm(\frac{3}{2} \left(1+\sqrt{2}\right))\}$+$\{(1,1,2+\sqrt{2})_{B_3}\pm(\frac{1}{2}+\sqrt{2})\}$+$\{(1,2,1+\sqrt{2})_{B_3}\pm(\frac{1}{2}+\frac{1}{\sqrt{2}})\}$+$\{(1+\sqrt{2},1,1+\sqrt{2})_{B_3}\pm(\frac{1}{2})\}$+$\{(2+\sqrt{2},1,1)_{B_3}\pm(0)\}$. \\
\noindent
\\ $(1,0,0,0)_{F_4}$= \newline
\noindent
\\$\{(0,0,0)_{B_3}\pm(1)\}$+$\{(0,0,\sqrt{2})_{B_3}\pm(0)\}$+$\{(1,0,0)_{B_3}\pm(\frac{1}{2})\}$. \\
\noindent
\\ $(1,0,0,1)_{F_4}$= \newline
\noindent
\\$\{(0,0,1)_{B_3}\pm(1+\frac{1}{\sqrt{2}})\}$+$\{(0,0,1+\sqrt{2})_{B_3}\pm(\frac{1}{\sqrt{2}})\}$+$\{(0,1,\sqrt{2})_{B_3}\pm(0)\}$+$\{(1,0,1)_{B_3}\pm(\frac{1}{2}+\frac{1}{\sqrt{2}})\}$+$\{(1,1,0)_{B_3}\pm(\frac{1}{2})\}$. \\
\noindent
\\ $(1,0,1,0)_{F_4}$= \newline
\noindent
\\$\{(0,1,0)_{B_3}\pm(1+\sqrt{2})\}$+$\{(0,1,1+\sqrt{2})_{B_3}\pm(\frac{1}{\sqrt{2}})\}$+$\{(1,1,0)_{B_3}\pm(\frac{1}{2}+\sqrt{2})\}$+$\{(1,1,1)_{B_3}\pm(\frac{1}{2}+\frac{1}{\sqrt{2}})\}$+$\{(\sqrt{2},0,1+\sqrt{2})_{B_3}\pm(0)\}$+$\{(1+\sqrt{2},0,1)_{B_3}\pm(\frac{1}{2})\}$. \\
\noindent
\\ $(1,0,1,1)_{F_4}$= \newline
\noindent
\\$\{(0,1,1)_{B_3}\pm(1+\frac{3}{\sqrt{2}})\}$+$\{(0,1,2+\sqrt{2})_{B_3}\pm(\sqrt{2})\}$+$\{(0,2,1+\sqrt{2})_{B_3}\pm(\frac{1}{\sqrt{2}})\}$+$\{(1,1,1)_{B_3}\pm(\frac{1}{2}+\frac{3}{\sqrt{2}})\}$+$\{(1,1,2)_{B_3}\pm(\frac{1}{2}+\sqrt{2})\}$+$\{(1,2,1)_{B_3}\pm(\frac{1}{2}+\frac{1}{\sqrt{2}})\}$+$\{(\sqrt{2},1,1+\sqrt{2})_{B_3}\pm(0)\}$+$\{(1+\sqrt{2},1,1)_{B_3}\pm(\frac{1}{2})\}$. \\  \newpage
\noindent
\\ $(1,1,0,0)_{F_4}$= \newline
\noindent
\\$\{(1,0,0)_{B_3}\pm(\frac{5}{2})\}$+$\{(1,0,2 \sqrt{2})_{B_3}\pm(\frac{1}{2})\}$+$\{(1,\sqrt{2},0)_{B_3}\pm(\frac{3}{2})\}$+$\{(2,0,0)_{B_3}\pm(2)\}$+$\{(2,0,\sqrt{2})_{B_3}\pm(1)\}$+$\{(3,0,0)_{B_3}\pm(\frac{1}{2})\}$. \\
\noindent
\\ $(1,1,0,1)_{F_4}$= \newline
\noindent
\\$\{(1,0,1)_{B_3}\pm(\frac{5}{2}+\frac{1}{\sqrt{2}})\}$+$\{(1,0,1+2 \sqrt{2})_{B_3}\pm(\frac{1}{2}+\frac{1}{\sqrt{2}})\}$+$\{(1,1,2 \sqrt{2})\}_{B_3}\pm(\frac{1}{2})\}$+$\{(1,\sqrt{2},1)_{B_3}\pm(\frac{3}{2}+\frac{1}{\sqrt{2}})\}$+$\{(1,1+\sqrt{2},0)_{B_3}\pm(\frac{3}{2})\}$+$\{(2,0,1)_{B_3}\pm(2+\frac{1}{\sqrt{2}})\}$+$\{(2,0,1+\sqrt{2})_{B_3}\pm(1+\frac{1}{\sqrt{2}})\}$+$\{(2,1,\sqrt{2})_{B_3}\pm(1)\}$+$\{(3,1,0)_{B_3}\pm(\frac{1}{2})\}$. \\
\noindent
\\ $(1,1,1,0)_{F_4}$= \newline
\noindent
\\$\{(1,1,0)_{B_3}\pm(\frac{5}{2}+\sqrt{2})\}$+$\{(1,1,1+2 \sqrt{2})_{B_3}\pm(\frac{1}{2}+\frac{1}{\sqrt{2}})\}$+$\{(1,1+\sqrt{2},0)_{B_3}\pm(\frac{3}{2}+\sqrt{2})\}$+$\{(1,1+\sqrt{2},1)_{B_3}\pm(\frac{3}{2}+\frac{1}{\sqrt{2}})\}$+$\{(2,1,0)_{B_3}\pm(2+\sqrt{2})\}$+$\{(2,1,1+\sqrt{2})_{B_3}\pm(1+\frac{1}{\sqrt{2}})\}$+$\{(1+\sqrt{2},0,1+2 \sqrt{2})_{B_3}\pm(\frac{1}{2})\}$+$\{(2+\sqrt{2},0,1+\sqrt{2})_{B_3}\pm(1)\}$+$\{(3+\sqrt{2},0,1)_{B_3}\pm(\frac{1}{2})\}$. \\
\noindent
\\ $(1,1,1,1)_{F_4}$= \newline
\noindent
\\$\{(1,1,1)_{B_3}\pm(\frac{5}{2}+\frac{3}{\sqrt{2}})\}$+$\{(1,1,2 \left(1+\sqrt{2}\right))_{B_3}\pm(\frac{1}{2}+\sqrt{2})\}$+$\{(1,2,1+2 \sqrt{2})_{B_3}\pm(\frac{1}{2}+\frac{1}{\sqrt{2}})\}$+$\{(1,1+\sqrt{2},1)_{B_3}\pm(\frac{3}{2} \left(1+\sqrt{2}\right))\}$+$\{(1,1+\sqrt{2},2)_{B_3}\pm(\frac{3}{2}+\sqrt{2})\}$+$\{(1,2+\sqrt{2},1)_{B_3}\pm(\frac{3}{2}+\frac{1}{\sqrt{2}})\}$+$\{(2,1,1)_{B_3}\pm(2+\frac{3}{\sqrt{2}})\}$+$\{(2,1,2+\sqrt{2})_{B_3}\pm(1+\sqrt{2})\}$+$\{(2,2,1+\sqrt{2})_{B_3}\pm(1+\frac{1}{\sqrt{2}})\}$+$\{(1+\sqrt{2},1,1+2 \sqrt{2})_{B_3}\pm(\frac{1}{2})\}$+$\{(2+\sqrt{2},1,1+\sqrt{2})_{B_3}\pm(1)\}$+$\{(3+\sqrt{2},1,1)_{B_3}\pm(\frac{1}{2})\}$. \\

\end{document}